\documentclass[aps,prl,preprint,superscriptaddress]{revtex4-1}

\usepackage{graphicx}
\usepackage{dcolumn}
\usepackage{bm}
\usepackage[mathlines]{lineno}
\usepackage{subfigure}

\begin{document}


\title{Measurement of the $CP$-violation Parameter sin2$\phi_1$ \\
with a New Tagging Method at the $\Upsilon(5S)$ Resonance}


\affiliation{University of Bonn, Bonn}
\affiliation{Budker Institute of Nuclear Physics SB RAS and Novosibirsk State University, Novosibirsk 630090}
\affiliation{Faculty of Mathematics and Physics, Charles University, Prague}
\affiliation{University of Cincinnati, Cincinnati, Ohio 45221}
\affiliation{Justus-Liebig-Universit\"at Gie\ss{}en, Gie\ss{}en}
\affiliation{Gifu University, Gifu}
\affiliation{Gyeongsang National University, Chinju}
\affiliation{Hanyang University, Seoul}
\affiliation{University of Hawaii, Honolulu, Hawaii 96822}
\affiliation{High Energy Accelerator Research Organization (KEK), Tsukuba}
\affiliation{Indian Institute of Technology Guwahati, Guwahati}
\affiliation{Institute of High Energy Physics, Chinese Academy of Sciences, Beijing}
\affiliation{Institute of High Energy Physics, Vienna}
\affiliation{Institute of High Energy Physics, Protvino}
\affiliation{Institute for Theoretical and Experimental Physics, Moscow}
\affiliation{J. Stefan Institute, Ljubljana}
\affiliation{Kanagawa University, Yokohama}
\affiliation{Institut f\"ur Experimentelle Kernphysik, Karlsruher Institut f\"ur Technologie, Karlsruhe}
\affiliation{Korea Institute of Science and Technology Information, Daejeon}
\affiliation{Korea University, Seoul}
\affiliation{Kyungpook National University, Taegu}
\affiliation{\'Ecole Polytechnique F\'ed\'erale de Lausanne (EPFL), Lausanne}
\affiliation{Faculty of Mathematics and Physics, University of Ljubljana, Ljubljana}
\affiliation{Luther College, Decorah, Iowa 52101}
\affiliation{University of Maribor, Maribor}
\affiliation{Max-Planck-Institut f\"ur Physik, M\"unchen}
\affiliation{University of Melbourne, School of Physics, Victoria 3010}
\affiliation{Graduate School of Science, Nagoya University, Nagoya}
\affiliation{Kobayashi-Maskawa Institute, Nagoya University, Nagoya}
\affiliation{Nara Women's University, Nara}
\affiliation{National United University, Miao Li}
\affiliation{Department of Physics, National Taiwan University, Taipei}
\affiliation{H. Niewodniczanski Institute of Nuclear Physics, Krakow}
\affiliation{Nippon Dental University, Niigata}
\affiliation{Niigata University, Niigata}
\affiliation{University of Nova Gorica, Nova Gorica}
\affiliation{Osaka City University, Osaka}
\affiliation{Pacific Northwest National Laboratory, Richland, Washington 99352}
\affiliation{Panjab University, Chandigarh}
\affiliation{Research Center for Nuclear Physics, Osaka University, Osaka}
\affiliation{RIKEN BNL Research Center, Upton, New York 11973}
\affiliation{University of Science and Technology of China, Hefei}
\affiliation{Seoul National University, Seoul}
\affiliation{Sungkyunkwan University, Suwon}
\affiliation{School of Physics, University of Sydney, NSW 2006}
\affiliation{Tata Institute of Fundamental Research, Mumbai}
\affiliation{Excellence Cluster Universe, Technische Universit\"at M\"unchen, Garching}
\affiliation{Toho University, Funabashi}
\affiliation{Tohoku Gakuin University, Tagajo}
\affiliation{Tohoku University, Sendai}
\affiliation{Department of Physics, University of Tokyo, Tokyo}
\affiliation{Tokyo Institute of Technology, Tokyo}
\affiliation{Tokyo Metropolitan University, Tokyo}
\affiliation{Tokyo University of Agriculture and Technology, Tokyo}
\affiliation{CNP, Virginia Polytechnic Institute and State University, Blacksburg, Virginia 24061}
\affiliation{Yamagata University, Yamagata}
\affiliation{Yonsei University, Seoul}
  \author{Y.~Sato}\affiliation{Tohoku University, Sendai} 
  \author{H.~Yamamoto}\affiliation{Tohoku University, Sendai} 
  \author{H.~Aihara}\affiliation{Department of Physics, University of Tokyo, Tokyo} 
  \author{D.~M.~Asner}\affiliation{Pacific Northwest National Laboratory, Richland, Washington 99352} 
  \author{V.~Aulchenko}\affiliation{Budker Institute of Nuclear Physics SB RAS and Novosibirsk State University, Novosibirsk 630090} 
  \author{T.~Aushev}\affiliation{Institute for Theoretical and Experimental Physics, Moscow} 
  \author{T.~Aziz}\affiliation{Tata Institute of Fundamental Research, Mumbai} 
  \author{A.~M.~Bakich}\affiliation{School of Physics, University of Sydney, NSW 2006} 
  \author{V.~Bhardwaj}\affiliation{Panjab University, Chandigarh} 
  \author{B.~Bhuyan}\affiliation{Indian Institute of Technology Guwahati, Guwahati} 
% ...
\collaboration{The Belle Collaboration}

\collaboration{Belle Collaboration}

\date{\today}

\begin{abstract}
We report a measurement of the $CP$-violation parameter sin2$\phi_1$
at the $\Upsilon(5S)$ resonance using a new tagging method,
called ``$B$-$\pi$ tagging.''
In $\Upsilon(5S)$ decays containing a neutral $B$ meson, a charged $B$, and a charged pion, the neutral $B$ is reconstructed in the $J/\psi K_S^0$ $CP$-eigenstate decay channel. 
The initial flavor of the neutral $B$ meson at the moment of the $\Upsilon(5S)$ decay is opposite to that of the charged $B$ and may thus  be inferred from the charge of the pion without reconstructing the charged $B$.
From the asymmetry between $B$-$\pi^+$ and $B$-$\pi^-$
tagged $J/\psi K_S^0$ yields,
we determine sin2$\phi_1$ = 0.57 $\pm$ 0.58(stat) $\pm$ 0.06(syst).
The results are based on 121 fb$^{-1}$ of data recorded
by the Belle detector at the KEKB $e^+ e^-$ collider.
\end{abstract}


\maketitle

In the Standard Model (SM),
$CP$-violation arises from an irreducible complex phase in the
Cabibbo-Kobayashi-Maskawa (CKM) quark mixing matrix \cite{CKM}.
Of the three angles of the unitary triangle,
$\phi_1 = {\rm arg} (-V_{cd} V_{cb}^*/V_{td} V_{tb}^*)$ \cite{phi1}
has been the most accessible, 
using the $B \rightarrow (c \bar{c}) K^0$ process,
because the hadronic uncertainty in this case is negligibly small.
$CP$-violation in the neutral $B$ meson system was clearly observed
and $\phi_1$ was measured
by the Belle \cite{phi1_belle} and
\mbox{\sl
B\hspace{-0.4em} {\small\sl A}\hspace{-0.37em} \sl B\hspace{-0.4em}
{\small\sl A\hspace{-0.02em}R}}
\cite{phi1_babar} collaborations.
These measurements used $B^0 \bar{B}^0$ pairs
that were produced at the $\Upsilon(4S)$ resonance;
the pairs are produced in a state with $C = -1$, where $C$ denotes the eigenvalue of charge conjugation.
Since the two $B$ mesons in the $C$-odd pair state are not allowed to have
the same $b$-flavor, $B^0 B^0$ or $\bar{B}^0 \bar{B}^0$,
the flavor of one $B$ meson is the opposite of the other $B$.
The other $B$ flavor is identified by combining information
from primary and secondary leptons, $K^{\pm}$,
$\Lambda$ baryons, slow and fast pions \cite{flavor-tagging}.
The mixing-induced $CP$-violation at the $\Upsilon$(4S)
vanishes in the time-integrated rates,
and thus a precise measurement of the distance
between the decay vertices of the two $B$ mesons is required.

The $CP$-violation parameter sin2$\phi_1$ can also be measured
at the $\Upsilon(5S)$ resonance using a new tagging method
which we call ``$B$-$\pi$ tagging'' \cite{bpi-tagging}.
In the decay of the $\Upsilon(5S)$ to $\bar B^{(*)0}B^{(*)+}\pi^-$ or its charge conjugate, the initial flavor of the neutral $B$ meson
is determined from the charge of the pion.
In the $B$-$\pi$ tagging method,
the neutral $B$ is fully reconstructed in a $CP$-eigenstate,
while the charged $B$ is not explicitly reconstructed
and identified indirectly through the recoil mass of the neutral $B$ 
and the charged pion.
This method works as well for events containing $B^*\to B\gamma$,
where one or more photons are present but not reconstructed.
The $CP$-violation parameter sin$2\phi_1$ can be obtained
from the time-integrated asymmetry of $BB\pi^+$ and $BB\pi^-$ tagged events:
\begin{eqnarray}
A_{BB\pi} &\equiv& \frac{N_{BB\pi^-} - N_{BB\pi^+}}{N_{BB\pi^-} + N_{BB\pi^+}}
\nonumber \\
&=&
\frac{{\cal S} x + {\cal A}}{1+x^2}
\label{eq:SA}
\end{eqnarray}
where $N_{BB\pi^{+}}$ and $N_{BB\pi^{-}}$ are
the observed number of $B^{(\ast)0} B^{(\ast)-} \pi^{+}$
and $\bar{B}^{(\ast)0} B^{(\ast)+} \pi^{-}$ events
in which the neutral $B$ decays to a $CP$-eigenstate, respectively
and ${\cal S}$ and ${\cal A}$ are the mixing-induced and direct $CP$-violating parameters, respectively.
The mixing parameter $x = (m_H - m_L) / \Gamma$,
where $\Gamma = (\Gamma_H + \Gamma_L)/2$,
is defined in terms of the masses $m_{H,L}$
and the decay widths $\Gamma_{H,L}$ of the heavy ($H$) and light ($L$)
neutral $B$ mass eigenstates.
The mixing parameter $y =(\Gamma_L - \Gamma_H)/2\Gamma$
is assumed to be zero,
as the SM predicts its value
to be negligibly small \cite{mixing_y_cal}
and the observed upper limit is ${\cal O}(10^{-2})$ \cite{mixing_y}.
In the case of $B \rightarrow (c\overline{c}) K^0_S$,
the SM predicts
${\cal S}=-\eta_{CP} {\rm sin}2\phi_1$
and ${\cal A}=0$
with very small theoretical uncertainty \cite{JpsiKstheory},
where $\eta_{CP}$ is the $CP$-eigenvalue of the final state.
Therefore, we can write
\begin{equation}
{\rm sin}2\phi_1 = -\eta_{CP} \left( \frac{1+x^2}{x} \right) A_{BB\pi}.
\end{equation}
There are several notable advantages to the $B$-$\pi$ tagging method.
First, $CP$-violation is observed through an asymmetry in event yields;
a measurement of the decay time of $B$ mesons is not required,
and associated systematic uncertainties are avoided.
Moreover, the method is applicable to decay channels such as $B \rightarrow \pi^0 \pi^0$,
in which it is difficult to measure decay vertices.
Current analyses at the $\Upsilon(4S)$ resonance
constrain only ${\cal A}$ for this mode \cite{pi0pi0}.
The analysis using this new tagging method
can give a constraint on the combination of the parameters ${\cal S}$ and ${\cal A}$.
In addition, as only one $B$ in the incoherent $B\bar{B}$ pair is reconstructed per event, systematic uncertainties associated with flavor tagging,
such as tag-side interference \cite{TSI}, do not arise. 
Finally, the $B$-$\pi$ tagging method can be extended
to higher $\Upsilon$ resonance decays.
For example, final states such as $\bar{B}_s^{(*)} B^{(*)+} K^{-}$
can be used to measure $CP$-violation in the $\bar B_s$ system
by tagging with a $K^{-}$.
Although the production cross section is smaller than
that at the $\Upsilon(4S)$ resonance, $B$-$\pi$ tagging at and above the $\Upsilon$(5S) is likely to become a powerful technique
at upgraded $B$-factories in the future.

In this Letter, we first measure the time-integrated mixing probability $\chi_d$ using the flavor specific modes $B^0 \rightarrow J/\psi K^{*0}$ and $D^{*-} \pi^+$ \cite{conjugate} to validate the $B$-$\pi$ tagging method.
We also measure direct $CP$-violation
in the charged $B^+ \rightarrow J/\psi K^+$ mode,
where the $CP$ asymmetry is known to be very small \cite{JpsichargedK}.
Finally, we report a measurement of sin2$\phi_1$ using the $CP$-eigenstate mode $B^0 \rightarrow J/\psi K_S^0$ with $\eta_{CP} = -1$.

The results reported here are based on 121 fb$^{-1}$ of data recorded
by the Belle detector \cite{BELLE} at the KEKB $e^+ e^-$ collider \cite{KEKB},
running at the center-of-mass (c.m.) energy of the $\Upsilon(5S)$ resonance.
The Belle detector is a general-purpose magnetic spectrometer
which consists of
a silicon vertex detector (SVD),
a 50-layer central drift chamber (CDC),
an array of aerogel threshold Cherenkov counters (ACC),
time-of-flight scintillation counters (TOF),
and an electromagnetic calorimeter (ECL) comprised of CsI(Tl) crystals.
The devices are located inside a superconducting solenoid coil
that provides a 1.5 T magnetic field.
An iron flux-return located outside the coil
is instrumented to detect $K_L^0$ mesons
and to identify muons (KLM).

All charged tracks other than $K_S^0 \rightarrow \pi^+ \pi^-$ daughters are required to originate from
the interaction point (IP).
Charged kaons and pions are identified by combining information from
the energy loss measurement in the CDC, 
the flight time measured by the TOF,
and the response of the ACC \cite{PID}.
Electrons are identified
by a combination of the energy loss measurement in the CDC,
the ratio of the cluster energy in the ECL
to the track momentum measured by the SVD and CDC,
and the shower shape in the ECL.
Muons are identified
by the track penetration depth and hit scatter in the KLM.

We reconstruct $J/\psi$ mesons
in the leptonic channels $e^+e^-$ or $\mu^+\mu^-$.
For the $e^\pm$ candidates,
we add the four-momentum of every photon detected
within 0.05 radians of the original track direction.
The invariant mass of $e^+ e^-$ pairs is then
required to satisfy
$-100$~MeV/$c^2<M_{e^+(\gamma)e^-(\gamma)}$-$m_{J/\psi}<+30$ MeV/$c^2$,
where $m_{J/\psi}$ is the nominal $J/\psi$ mass;
the interval is asymmetric
because small residual radiative tails remain.
For $\mu^+ \mu^-$ pairs, we require the invariant mass to be within 30 MeV/$c^2$ of the nominal $J/\psi$ mass.
The $J/\psi$ mass resolution is about 10 MeV/$c^2$.
 
The $K_S^0$ candidates are formed
by combining two oppositely charged tracks,
assuming both are pions.
Since the $K_S^0$'s can be selected with low background,
we apply a loose mass selection that requires an invariant mass
within 30 MeV/$c^2$ of the $K^0$ mass.
We then impose the following additional requirements:
(1) the two pion tracks must have a large distance of closest approach to the IP
in the plane perpendicular to the electron beam line;
(2) the pion tracks must intersect at a common vertex
that is displaced from the IP;
(3) the $K_S^0$ candidate's momentum vector should originate from the IP.

Candidates for $K^{*0}$ and $\bar{D}^0$ mesons are reconstructed
in the $K^{*0} \rightarrow K^+ \pi^-$
and $\bar{D}^0 \rightarrow K^+ \pi^-$ channels, respectively.
They are formed by combining oppositely charged kaon
and pion tracks and requiring the invariant mass to lie
within 50 MeV/$c^2$($\sim 1 \Gamma$) for $K^{*0}$
and within 10 MeV/$c^2$($\sim 2 \sigma$) for $\bar{D}^0$
of the nominal masses, respectively.
$D^{*-}$ candidates are reconstructed by combining
a $\bar{D}^0$ candidate with a $\pi^-$.
The mass difference between
the $D^{*-}$ and $\bar{D}^0$ candidates
is then required to be within 2 MeV/$c^2$($\sim 3.5 \sigma$)
of the nominal mass difference.

The $B$ candidates are required to have an invariant mass
within 20 MeV/$c^2$ of the $B$ mass,
which corresponds to approximately $\pm 2\sigma$,
$\pm 2.7\sigma$, 
$\pm 2.4\sigma$ and 
$\pm 3\sigma$ intervals
for the $D^{*-} \pi^+$, $J/\psi K^{*0}$,
$J/\psi K^+$, and $J/\psi K_S^0$ modes, respectively.
To select $B$ mesons in $\Upsilon(5S)\to B^{(*)}\bar B^{(*)}\pi$ events,
we require 5.348 GeV/$c^2$ $<$ $M_{\rm bc}$ $<$ 5.440 GeV/$c^2$,
where $M_{\rm bc}$ is the beam-energy-constrained mass,
$M_{\rm bc} = \sqrt{(E_{\rm beam}^{\rm cms})^2 - (p_{B}^{\rm cms})^2}$.
The quantities $E_{\rm beam}^{\rm cms}$ and $p_{B}^{\rm cms}$ are
the beam energy and momentum of the $B$ candidate
in the c.m.\ frame.
If we neglect detector resolution,
$M_{\rm bc}$ is less than 5.325 GeV/c$^2$
in $\Upsilon(5S) \to B^{(*)} B^{(*)}$ events.
On the other hand,
$M_{\rm bc}$ is higher than 5.351 GeV/c$^2$
in $\Upsilon(5S) \to B^{(*)} B^{(*)}\pi$ events.
Even if we consider the effect of detector resolution,
we can separate $\Upsilon(5S) \to B^{(*)} B^{(*)}\pi$ and
$\Upsilon(5S) \to B^{(*)} B^{(*)}$ decays.

To suppress $e^+e^- \rightarrow q \overline{q}$ $(q=u,d,s,c)$ continuum backgrounds,
we apply selections on topological variables measured in the c.m. system.
The ratio of the second to the zeroth Fox-Wolfram moments \cite{Fox}
is required to be less than
0.5 for the $J/\psi$ final states,
$J/\psi K^{*0}$, $J/\psi K^+$ and $J/\psi K_S^0$,
for which the background level is low,
and less than 0.4 for the $D^{*-} \pi^+$ final state.
To further reduce the continuum background for the $D^{*-} \pi^+$ mode,
the angle between the thrust axis of the particles
forming the $B$ candidate and the thrust axis of all other particles
in the event, $\theta_{\rm thr}^{\rm cms}$, is required to satisfy
$|{\rm cos}\theta_{\rm thr}^{\rm cms}| < 0.75$.
These selections retain 98\% (78\%) of the signal
and remove 18\% (74\%) of the background
for the $J/\psi$ ($D^{*-}\pi^+$) final states.
More than one $B$ candidate per event is allowed.
The probability of multiple candidates, however, is less than 1\% per event.
The reconstruction efficiencies
for the $D^{*-} \pi^+$, $J/\psi K^{*0}$,
$J/\psi K^+$, and $J/\psi K_S^0$ modes
are 24.6\%, 21.2\%, 44.4\%, and 37.7\%, respectively.

We then combine a reconstructed $B$ candidate with each charged pion
that was not used in the reconstruction of the $B$ candidate.
The point of closest approach for the pion is required to be within 1~cm of the vertex of the reconstructed $B$ decay in the plane perpendicular to the electron beam.
Pion tracks identified as $K_S^0\to\pi^+\pi^-$ daughters are
rejected.
We calculate the missing mass:
\begin{eqnarray}
M_{\rm miss}^2 \equiv 
\left[ P_{\rm total} - 
\left(
P_{B} + P_{\pi^{\pm}}
\right)
\right]^2
\end{eqnarray}
where $P_{\rm total}$, $P_{B}$, and $P_{\pi^{\pm}}$ are the
total 4-momenta of the initial state, the reconstructed $B$ meson candidate,
and the pion candidate, respectively. 
To improve the missing mass resolution, mass- and vertex-constrained fits are applied to
$B$, $J/\psi$, $K_S^0$, $D^{*+}$ and $D^0$ candidates,
and a vertex-constrained fit is applied to $K^{*}$ candidates.
The tagging efficiencies
for the $B\bar{B}\pi$, $B^*\bar{B}\pi+B\bar{B}^* \pi$,
and $B^*\bar{B}^*\pi$ decay channels
are 70.6\%, 64.7\%, and 54.1\%, respectively.
For the $BB\pi$ channel, the missing mass is equal to the nominal $B$ mass,
while for the $B \bar{B}^* \pi$, $B^* \bar{B} \pi$
and $B^* \bar{B}^* \pi$ channels,
the missing mass is shifted by approximately the $B^*$-$B$ mass difference (46~MeV)
for each unreconstructed photon from a $B^*$ decay.

\begin{figure}[hb]
{\includegraphics[width=7.5cm]{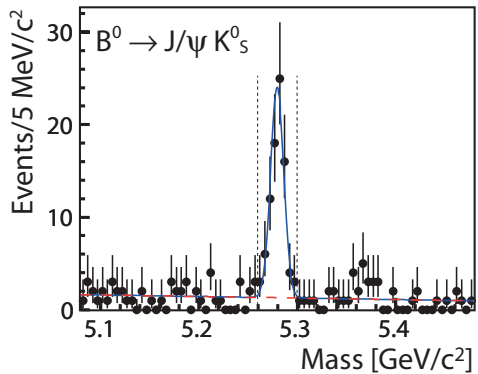}}
\caption{
Invariant mass of reconstructed $B^0 \rightarrow J/\psi K_S^0$ candidates.
The background component is shown by the dashed curve.
The sum of signal and background components is shown by the solid curve.
The vertical lines show the requirement on the $B^0$ mass.
\label{fig:recb_jpsiks}
}
\end{figure}
\begin{figure}[ht]
\subfigure
{\includegraphics[width=7.5cm]{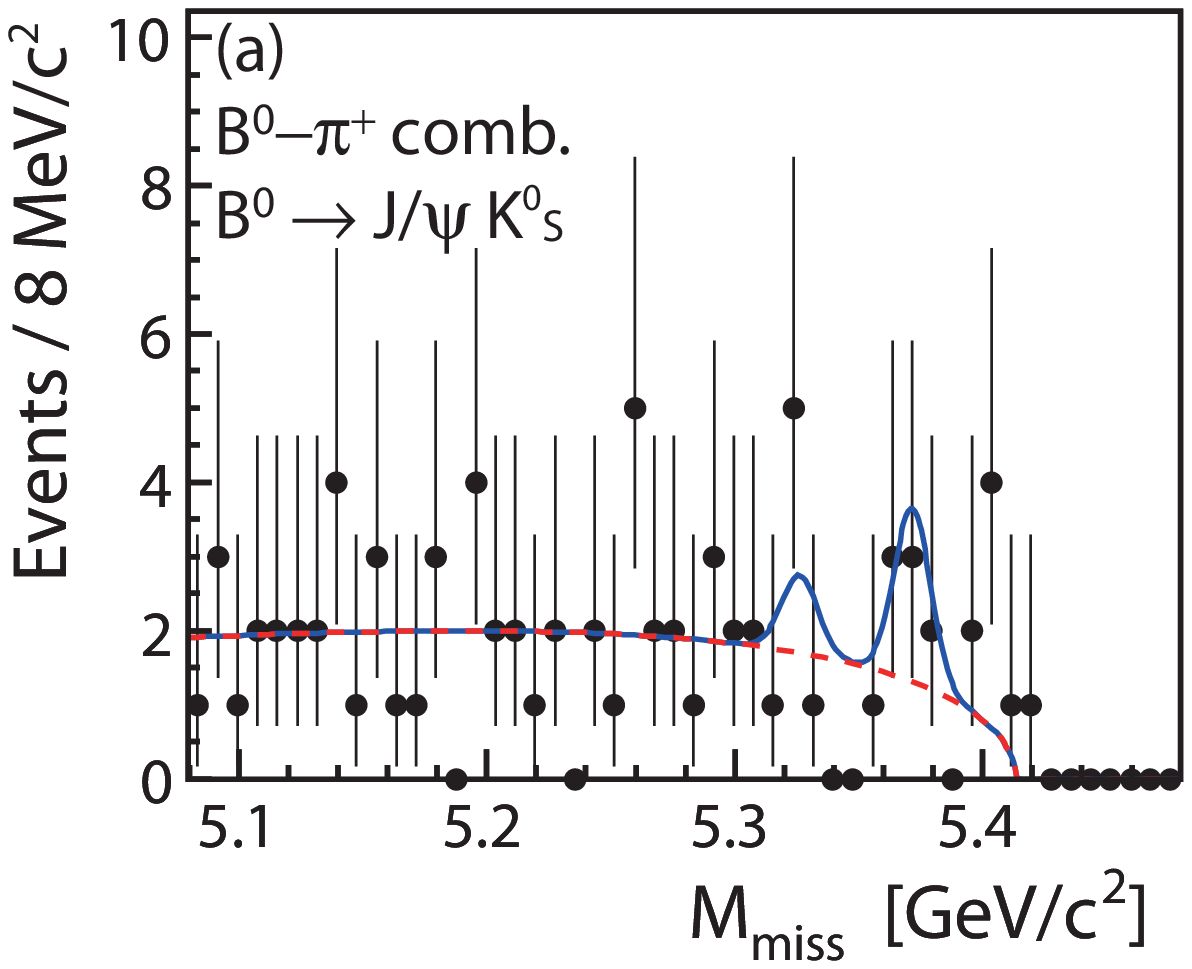}}
\subfigure
{\includegraphics[width=7.5cm]{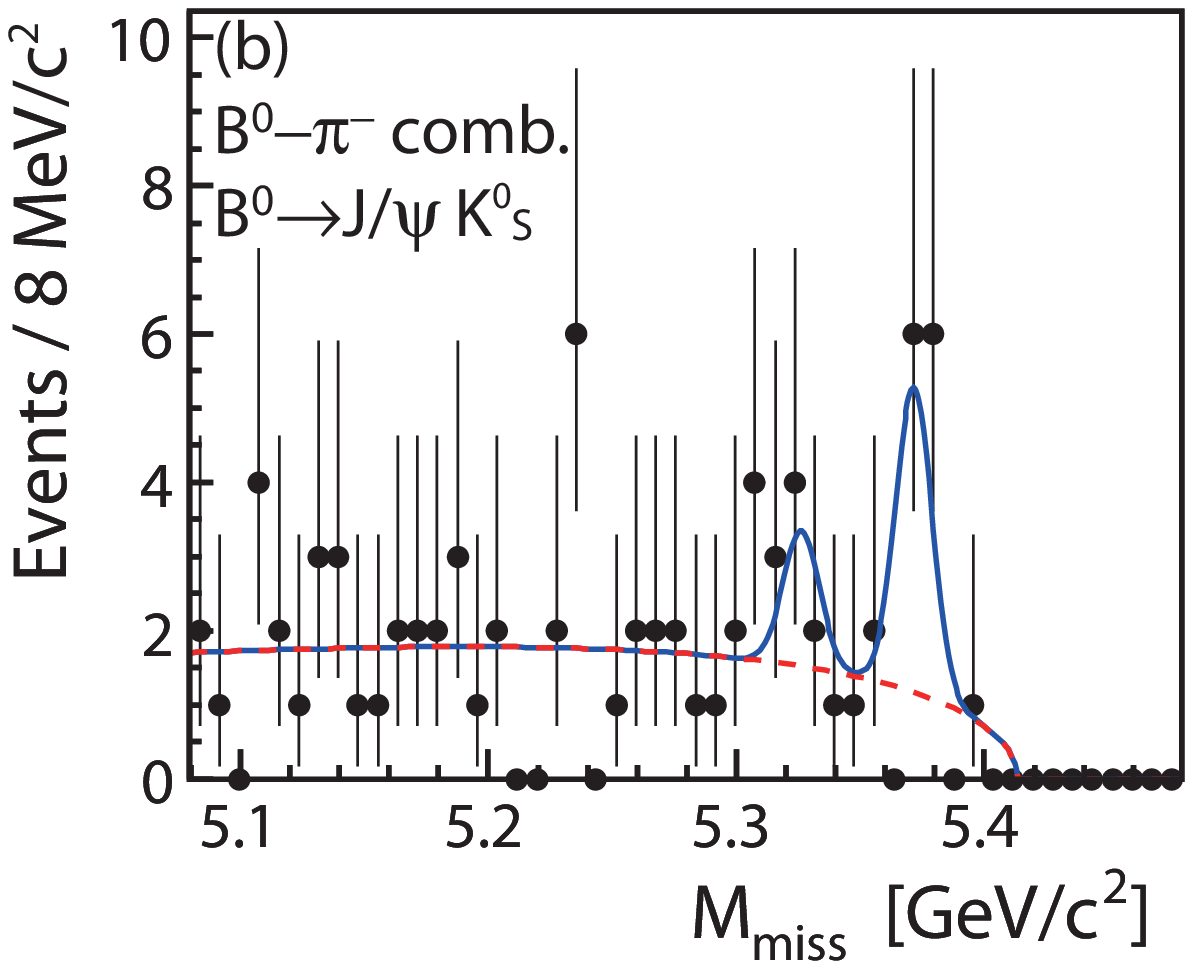}}
\caption{
Missing mass distributions for $B^0 \rightarrow J/\psi K_S^0$ candidates
tagged by (a) $\pi^+$ and (b) $\pi^-$ in the $\Upsilon(5S)$ data sample.
The solid curve is the fit projection for the sum of signal and background.
The dashed curve shows the background component.
Two peaks correspond to
the $B\bar{B}^* \pi + B^* \bar{B}\pi$ (left)
and $B^*\bar{B}^*\pi$ (right) decay channels, respectively.
\label{fig:misb_jpsiks}
}
\end{figure}

The dominant sources of background are random track combinations
and initial-state radiation (ISR) processes
i.e. $e^+e^- \rightarrow \Upsilon(4S) \gamma \rightarrow B \bar{B} \gamma$ \cite{BBP}.
The combinatorial background arises mainly from
combinations of correctly reconstructed $B$ mesons with pions from the other $B$.
These backgrounds do not peak in the missing mass distribution.

To validate the $B$-$\pi$ tagging method,
we extract the time-integrated mixing probability,
$\chi_d$,
from the missing mass distribution
in the flavor-specific modes
$B \rightarrow J/\psi K^{*0}$ and $D^{*-} \pi^+$.
For each mode, events are sorted into four categories
according to the flavor of the $B$ candidate and the charge of the pion.
The $B^0$-$\pi^+$ and $\bar{B}^0$-$\pi^-$ combinations are unmixed
while $\bar{B}^0$-$\pi^+$ and $B^0$-$\pi^-$ are mixed processes.
The value of $\chi_d$ is extracted from
the yields of mixed and unmixed processes,
$N_{\rm mixed}$ and $N_{\rm unmixed}$, respectively:
\begin{equation}
\chi_d = \frac{N_{\rm mixed}}{N_{\rm mixed} + N_{\rm unmixed}}.
\end{equation}
An extended unbinned maximum likelihood fit in missing mass is
simultaneously applied to the $J/\psi K^{*0}$ and $D^{*-} \pi^+$ samples.
The $B \bar{B}^* \pi + B^* \bar{B} \pi$ and $B^* \bar{B}^* \pi$ signals
are modeled by two Gaussians
with parameters fixed from Monte Carlo (MC) samples. 
The ratio of the sum of $B \bar{B}^* \pi$ and $B^* \bar{B} \pi$
to the $B^* \bar{B}^* \pi$ yield is floated.
The $B \bar{B} \pi$ and $B \bar{B} \pi \pi$ decay channels
are not included in the fit,
as their contributions were found to be negligible
in other analyses \cite{BBP}.
The $BB\pi$ channel is expected to be suppressed by angular momentum considerations,
and the $BB\pi\pi$ channel is expected
to be suppressed due to the limited phase space.
The combinatorial and ISR backgrounds are
described by an ARGUS function \cite{ARGUS}.
The endpoint of the ARGUS function is fixed from the MC samples.
The background yields are floated independently in the fits
to the four $B$-$\pi$ charge combinations.
Assuming there is no direct $CP$-violation,
we obtain $\chi_d = 0.19 \pm 0.09({\rm stat})$,
which is consistent with the current world average
of $0.1864 \pm 0.0022$ \cite{HFAG}.
The $J/\psi K^{*0}$ and $D^{*-} \pi^+$ signal yields are
$41.2 \pm 9.5$ and $29.6 \pm 9.0$ events, respectively.

We also check direct $CP$-violation
from a charged $B$ mode, $B^+ \rightarrow J/\psi K^+$.
The missing mass distributions for
$B^+$-$\pi^-$ and $B^-$-$\pi^+$ combinations
are fitted with the same signal and background functions
as used for the flavor specific modes.
The signal yield is $64.8 \pm 11.9$ events.
We find $A_{BB\pi} = 0.02 \pm 0.17$,
which is consistent with zero asymmetry, as expected.
These two measurements validate the $B$-$\pi$ tagging method
within the available statistics.

\begin{figure}[ht]
{\includegraphics[width=8.0cm]{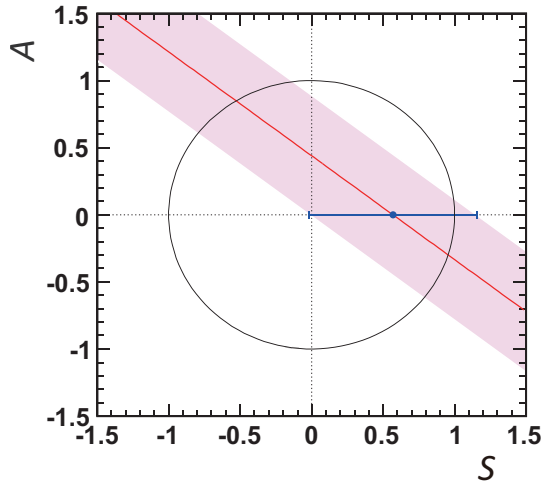}}
\caption{
Confidence region for ${\cal S}$ and ${\cal A}$.
The circle shows the physical boundary.
The shaded region shows the $\pm 1 \sigma$ region
using the $B$-$\pi$ tagging method at the $\Upsilon(5S)$ resonance
and the point with an error bar is the ${\cal S}={\rm sin}2\phi_1$ measurement
assuming no direct $CP$ violation (${\cal A}=0$).
\label{fig:sa_plot}
}
\end{figure}

The sin2$\phi_1$ parameter is extracted from the $CP$-eigenstate mode,
$B^0 \rightarrow J/\psi K_S^0$.
The $B$ meson yield is 75.9 $\pm$ 9.5 events
as determined from a fit to the reconstructed $B$ mass distribution
as shown in Fig.~\ref{fig:recb_jpsiks}.
The signal and background are fitted with a Gaussian and
a first-order Chebyshev polynomial, respectively. 
We then fit the $\pi^+$ and $\pi^-$ tagged samples
in the missing mass distributions
with the same signal and background functions
as used for the flavor specific and charged modes.
The result is shown in Fig.~\ref{fig:misb_jpsiks}.
Two peaks correspond to the $B\bar{B}^* \pi + B^* \bar{B}\pi$
and $B^*\bar{B}^*\pi$ decay channels, respectively.
We obtain $A_{BB\pi}=0.28 \pm 0.28({\rm stat})$.
The signal yields tagged by $\pi^+$ and $\pi^-$ mesons
are $7.8 \pm 3.9$ and $13.7 \pm 5.3$ events, respectively.
Figure~\ref{fig:sa_plot} shows
the resulting two-dimensional confidence regions
in the ${\cal S}$ and ${\cal A}$ plane
from Eq. (\ref{eq:SA}),
taking the mixing parameter $x$ to be $0.771 \pm 0.007$ \cite{HFAG}.
Assuming ${\cal A}=0$, we obtain
\begin{equation}
{\rm sin}2\phi_1 = 0.57 \pm 0.58 ({\rm stat}) \pm 0.06 ({\rm syst}).
\end{equation}

The dominant systematic uncertainty for sin2$\phi_1$ arises
from the signal and background shape parameters fixed with MC samples.
This uncertainty is evaluated by varying the fitted parameters,
the means and width of the two Gaussians for the signal
and the endpoint of the ARGUS function,
by the difference observed between the data and MC samples for $B \rightarrow J/\psi K^{*0}$ and $D^{*-} \pi^+$ and found to be 0.055.
The systematic uncertainty from possible $B \bar{B} \pi$ and $B \bar{B} \pi \pi$ contributions is estimated to be 0.005
by refitting the data using a fitting function that includes $B B \pi$ and $B B \pi \pi$.
The ratios of $B \bar{B} \pi$ and $B \bar{B} \pi \pi$
to the sum of $B \bar{B}^* \pi$ and $B^* \bar{B} \pi$ are set to the upper limits determined in the $B \rightarrow J/\psi K^{*0}$ and $D^{*-} \pi^+$ modes.
The systematic uncertainty from a possible pion reconstruction asymmetry
is evaluated to be 0.015
using the following equation:
\begin{equation}
\frac{\epsilon^{\pi^+}}{\epsilon^{\pi^-}}
=
\frac{N_{D^{*+}}/N_{D^0}}{N_{D^{*-}}/N_{\bar{D}^0}}
\end{equation}
where $\epsilon^{\pi^{\pm}}$ is the detection efficiency of $\pi^{\pm}$
and $N_{D^{*+}}$ $(N_{D^0})$ is the total number of reconstructed
$D^{*+}$ $(D^0)$ mesons in the $\Upsilon(4S)$ data sample. 
The $D^{*+}$ is reconstructed from $D^{0} \pi^+$,
and $D^0$ is reconstructed from $K^- \pi^+$.
We require pions from $D^{*+}$
to be in the kinematic region accessible to pions from $\Upsilon(5S)$ decays.
Since the detection efficiencies for kaons and pions from $D^0$ cancel, the detection efficiency for pions from the $D^{*+}$ decay can be evaluated.
The ratio $\epsilon^{\pi^+}/\epsilon^{\pi^-}$ is estimated to be $1.009 \pm 0.007$,
and $1.016$ is used for the calculation of the systematic uncertainty.
The systematic uncertainties from the mixing parameters $x$ and $y$ \cite{HFAG,mixing_y} are estimated to be  0.001 and 0.012, respectively.
The total systematic uncertainty is estimated by summing the above uncertainties in quadrature and found to be 0.058.

In conclusion, we report a measurement of sin2$\phi_1$ with a new tagging method called $B$-$\pi$ tagging, using a 121~fb$^{-1}$ data sample collected at the $\Upsilon(5S)$ resonance.
This method is complementary to time-dependent analyses using flavor tagging methods at the $\Upsilon(4S)$ resonance \cite{phi1_belle,phi1_babar}.
We measure sin2$\phi_1$ to be 0.57$\pm$0.58(stat)$\pm$0.06(syst),
which is consistent with the value
obtained on the $\Upsilon(4S)$ resonance.
The $B$-$\pi$ tagging method allows the measurement of $CP$-violation without decay time information and thus has great potential
for the $B \rightarrow \pi^0 \pi^0$
and $\bar{B}_s^{(*)} B^{(*)+} K^-$ decay channels
at future high luminosity $B$-factory experiments.
 
This work was supported in part by a Grant-in-Aid for JSPS Fellows, No.10J03308.
We thank the KEKB group for excellent operation of the
accelerator; the KEK cryogenics group for efficient solenoid
operations; and the KEK computer group, NII, and 
PNNL/EMSL for valuable computing and SINET4 for network support.  
We acknowledge support from MEXT, JSPS and Nagoya's TLPRC (Japan);
ARC and DIISR (Australia); NSFC (China); MSMT (Czechia);
DST (India); INFN (Italy); MEST, NRF, GSDC of KISTI, and WCU (Korea); 
MNiSW (Poland); MES and RFAAE (Russia); ARRS (Slovenia); 
SNSF (Switzerland); NSC and MOE (Taiwan); and DOE and NSF (USA).


\begin{thebibliography}{99}
\bibitem{conjugate}
Charge-conjugate modes are implied
through this Letter unless specified.
\end{thebibliography}
\end{document}